\documentclass[12pt]{article}
\usepackage{amsmath,amsfonts}
\usepackage[nosort]{cite}

\def\bR {\mathbb{R}}

\newcommand{\Tr}{{\rm Tr\,}}

\newcommand{\cN}{{\mathcal N}}
\newcommand{\cP}{{\mathcal P}}
\newcommand{\cQ}{{\mathcal Q}}

\renewcommand{\title}[1]{\vbox{\center\LARGE{#1}}\vspace{5mm}}
\renewcommand{\author}[1]{\vbox{\center#1}\vspace{5mm}}
\newcommand{\address}[1]{\vbox{\center\em#1}}
\newcommand{\email}[1]{\vbox{\center\tt#1}\vspace{5mm}}

\begin{document}
\begin{titlepage}
\begin{center}
\hfill {\tt HU-EP-07/13}\\
\hfill {\tt YITP-SB-07-12}\\
\hfill{\tt Imperial/TP/07/RR/02}\\
\hfill {\tt arXiv:0704.2237}\\
\vskip .5cm

\title{More supersymmetric Wilson loops}

\author{Nadav Drukker$^{1,a}$,
Simone Giombi$^{2,b}$,
Riccardo Ricci$^{3,4,c}$,
Diego Trancanelli$^{2,d}$}

\address{$^1$Humboldt-Universit\"at zu Berlin, Institut f\"ur Physik,\\
Newtonstra{\ss}e 15, D-12489 Berlin, Germany\\
\medskip
$^2$C. N. Yang Institute for Theoretical Physics,\\
State University of New York at Stony Brook\\
Stony Brook, NY 11794-3840, USA\\
\medskip
$^3$ Theoretical Physics Group, Blackett Laboratory,\\
Imperial College, London, SW7 2AZ, U.K.\\
\medskip
$^4$ The Institute for Mathematical Sciences, \\
Imperial College, London, SW7 2PG, U.K.}

\email{$^a$drukker@physik.hu-berlin.de,
$^b$sgiombi@max2.physics.sunysb.edu,
$^c$r.ricci@imperial.ac.uk,
$^d$dtrancan@max2.physics.sunysb.edu}

\end{center}

\abstract{
\noindent
We present a large new family of Wilson loop operators in $\cN=4$ 
supersymmetric Yang-Mills theory. For an arbitrary curve on the 
three dimensional sphere one can add certain scalar couplings to the 
Wilson loop so it preserves at least two supercharges. 
Some previously known loops, notably the $1/2$ BPS circle, belong to 
this class, but we point out many more special cases which were not 
known before and could provide further tests of the $AdS$/CFT 
correspondence.
}

\vfill

\end{titlepage}

Supersymmetry is an extremely powerful tool in theoretical physics. 
In addition 
to the general hope and expectation that supersymmetry will be discovered 
at high energies, it affords the theorist extra freedom and control. Field 
theories with supersymmetry show similar phenomena to non-supersymmetric 
theories but are often easier to work with. Particularly, supersymmetric 
theories have special operators which are invariant under some of the 
supersymmetry generators and therefore belong to shorter multiplets of the 
algebra and may be protected from quantum corrections.

In $\cN=4$ supersymmetric Yang-Mills theory the local operators preserving 
some of the supersymmetry generators are well studied. In the dual string theory 
on $AdS_5\times S^5$ they are described by the Kaluza-Klein modes 
arising in 
the reduction of the supergravity fields on $S^5$, or by giant gravitons. 
The supersymmetric non-local operators, such as 
Wilson loops, are not nearly as well understood. So far the only 
supersymmetric Wilson loops with a non-trivial expectation value that 
were described had a circular geometry.

In this letter we will present 
a new class of Wilson loop operators that preserve between 2 and 
16 supercharges. Since they are non-local they might capture interesting 
properties of this gauge theory. In the dual string theory they will 
be described by string surfaces (or D-branes) and due to their 
supersymmetry they may be under better computational control than 
non-supersymmetric Wilson loops. Indeed in a few cases their 
expectation values as calculated in the field theory and string theory 
provide an amazing test for the $AdS$/CFT correspondence.

We present here the main idea and several special examples of 
interesting loops that belong to this class. We will study those loops in 
greater detail in future publications \cite{next}.

In addition to the required coupling of the gauge field $A_\mu$ to the tangent 
vector $\dot x^\mu(s)$, it is natural to couple the Wilson loop in $\cN=4$ 
YM to the scalars $\Phi^I$ (with $I=1,\ldots,6$) by the functions 
$\Theta^I(s)$ so it takes the form \cite{Rey,Maldacena-wl}
\begin{equation}
W=\frac{1}{N}\Tr\,\cP\exp \oint ds 
\left(iA_\mu\dot x^\mu(s) + |\dot x|\Theta^I(s)\Phi^I\right)\,.
\label{Wilson-loop}
\end{equation}
Requiring that such a loop be supersymmetric leads to an equation at 
every point along the path $x^\mu(s)$. Only if all those 
conditions commute, will the loop be globally supersymmetric.

One simple way to satisfy this is if at every point one finds the same equation. 
This happens in the case of the straight line, where $\dot x^\mu$ is a 
constant vector and one takes also $\Theta^I$ to be a constant. This 
idea was generalized in a very ingenious way by Zarembo 
\cite{Zarembo:2002an}, who assigned to every tangent vector in 
$\bR^4$ a unit vector in $\bR^6$ (through a $6\times 4$ matrix $M^I{}_\mu$) 
and took $|\dot x|\Theta^I=M^I{}_\mu\dot x^\mu$. That construction 
guarantees that if a curve is contained within a one-dimensional linear 
subspace of $\bR^4$ it is $1/2$ BPS. Inside a 2-plane it will be $1/4$ BPS, 
inside $\bR^3$ it's $1/8$ BPS and a generic curve is $1/16$ BPS.

An amazing 
fact about those loops is that their expectation value seems to be trivial 
\cite{Guralnik:2003di,Guralnik:2004yc,Dymarsky:2006ve}. 
But this is also a crucial shortcoming of this construction, 
one of the most interesting Wilson loop observables is the circle with a 
coupling to a single scalar, whose expectation value is a non-trivial function 
of both the rank of the gauge group $N$ and of the 't Hooft coupling 
$\lambda=g_{YM}^2N$ \cite{Erickson:2000af,Drukker:2000rr}. 
This Wilson loop preserves $1/2$ of the 
supersymmetries, but is not given by the above construction. Recently 
some $1/4$ BPS loops were described that also do not have trivial 
expectation values, rather the values of all those loops seem to be 
described by a 0-dimensional matrix model 
\cite{Drukker:2006ga}.

The supersymmetries preserved by those loops with non-unit 
expectation values always include the superconformal generators 
(written usually as $S$ and $\bar S$) in addition to the usual 
Poincar\'e supercharges ($Q$ and $\bar Q$). The loops with trivial 
expectation values are annihilated by combinations of $Q$ and $\bar Q$ 
alone, so our construction below gives more loops annihilated by 
combinations including also $S$ and $\bar S$.

The family of Wilson loops that we present here follows an arbitrary 
curve on $S^3$, which may be the
unit sphere in flat $\bR^4$ in the Euclidean gauge theory, or a 
spatial slice for the Lorentzian theory on $S^3\times\bR$. The 
basic ingredient in our construction are 
the invariant one-forms on the group manifold $SU(2)=S^3$ 
(we follow the conventions of \cite{Gauntlett:1995fu}). 
In terms of flat coordinates $x^{\mu}$ satisfying $x^2=1$ they read 
\begin{equation}
\begin{aligned}
\sigma_1^{R,L}& = 2 \left[\pm(x^2 dx^3-x^3 dx^2) +
(x^4 dx^1-x^1 dx^4) \right], \\
\sigma_2^{R,L} &= 2 \left[\pm(x^3 dx^1-x^1 dx^3) + 
(x^4 dx^2-x^2 dx^4) \right], \\
\sigma_3^{R,L} &= 2 \left[\pm(x^1 dx^2-x^2 dx^1) + 
(x^4 dx^3-x^3 dx^4) \right],
\label{one-forms}
\end{aligned}
\end{equation}
where $\sigma_i^R$ are the right (or left-invariant) one-forms and 
$\sigma_i^L$ are the left (or right-invariant) one-forms. These are 
respectively dual to left (right) invariant vector fields $\xi_i^R$ 
($\xi_i^L$) generating right (left) group actions. We can now use 
either $\sigma_i^R$ or $\sigma_i^L$ to define a natural coupling 
to three of the scalars. We will choose to use the right one-forms. 
Our ansatz for the supersymmetric Wilson loop on $S^3$ is then 
\begin{equation}
W=\frac{1}{N}\Tr\,\cP\exp \oint  
\left( i A + \frac{1}{2} \sigma_i^R M^i{}_I\Phi^I \right) ,
\label{susy-loop}
\end{equation} 
where for convenience we wrote the integral in form notation. 
The $3\times 6$ matrix $M^i{}_I$ specifies which three scalars the 
loop will couple to and satisfies that 
$M M^{\top}$ is the $3\times 3$ unit matrix. 
When we need an explicit choice of $M$ we take 
$M^1{}_1=M^2{}_2=M^3{}_3=1$ and all other entries zero.

The supersymmetry variation of the Wilson loop will be proportional to
\begin{equation}
\delta W \simeq \left(i\,dx^\mu\gamma_\mu 
+\frac{1}{2}\sigma_i^R M^i{}_I\rho^I\gamma^5
\right) \epsilon(x)\,,
\label{susy}
\end{equation}
where the pullback of the one-forms along the curve is assumed. 
$\gamma_\mu$ and $\rho^I$ are respectively the gamma 
matrices of $SO(4)$ and $SO(6)$, the Poincar\'e and R-symmetry 
groups and they commute with each-other. 
$\gamma^5=-\gamma^1\gamma^2\gamma^3\gamma^4$ is the four 
dimensional chirality matrix and $\epsilon(x)$ is a 
conformal Killing spinor given by two arbitrary constant 
spinors (which are also spinors of the R-symmetry group) as
\begin{equation}
\epsilon=\epsilon_0 + x^\mu\gamma_\mu\epsilon_1\,.
\end{equation}
Note now that the action of the $\gamma$'s on the chiral components of a 
spinor $\epsilon^\pm=\frac{1}{2}(1\pm\gamma^5)\epsilon$ 
can be expressed in terms of the identity and Pauli matrices 
$\pm i\tau_i$, allowing one to write
\begin{equation}
i\,dx^\mu x^\nu\gamma_{\mu\nu}\epsilon^\mp
=\pm\frac{1}{2}\tau^i\sigma^{R,L}_i\epsilon^\mp\,.
\end{equation}
Therefore if we restrict to the anti-chiral components of $\epsilon$, 
equation (\ref{susy}) can be written as
\begin{equation}
\delta W \simeq \frac{1}{2}\,\sigma_i^R\left(
\tau^i\epsilon_1^--M^i{}_I\rho^I\epsilon_0^-
-x^\eta\gamma_\eta(\tau^i\epsilon_0^-
-M^i{}_I\rho^I\epsilon_1^-)
\right).
\end{equation}
This equation is solved if
\begin{equation}
\tau^i\epsilon^-_1=M^i{}_I\rho^I\epsilon^-_0\,.
\label{constraints}
\end{equation}
On the other hand, the chiral part of $\epsilon$ will 
introduce $\sigma^L_i$ into (\ref{susy}), so unless there are linear relations
between the six $\sigma^{R,L}_i$ and/or $x^\eta\gamma_\eta$, 
one finds that $\epsilon^+=0$. While this is true for a generic curve 
on $S^3$, some of the special examples we detail below are degenerate 
and posses extra supersymmetries.

To solve the set of equations (\ref{constraints}), let us choose the matrix 
$M$ that identifies $i$ with $I$. Then we can eliminate $\epsilon_0^-$ 
to get
\begin{equation}
i\tau_i\epsilon_1^-
=-\frac{1}{2}\epsilon_{ijk}\rho^{jk}\epsilon_1^-\,,\qquad
i=1,\,2,\,3\,.
\label{newconstraints}
\end{equation}
This is a consistent set of three constraints, out of which only two 
are independent. Since 
$\epsilon^-_1$ has eight real components and the other spinor, 
$\epsilon^-_0$, is determined from it, we conclude that for a generic curve on 
$S^3$ the Wilson loop preserves $1/16$ of the original supersymmetries. 

To find explicitly the combinations of $\bar Q$ and 
$\bar S$ which leave the Wilson loop invariant, notice that singling out 
three of the scalars breaks the R-symmetry group $SU(4)$ 
down to $SU(2)_A \times SU(2)_B$, where $SU(2)_A$ acts on 
$\Phi^1,\Phi^2,\Phi^3$ while $SU(2)_B$ rotates 
$\Phi^4,\Phi^5,\Phi^6$. The operators 
appearing in (\ref{newconstraints}) are just the generators of 
$SU(2)_R$ and $SU(2)_A$, and the above equations simply state 
that $\epsilon^-_1$ is a singlet of the diagonal sum of $SU(2)_R$ 
and $SU(2)_A$.
If we split the $SU(4)$ index $A$ into $\dot a$ for $SU(2)_A$ and $a$ 
for $SU(2)_B$, then the singlet can be written as 
\begin{equation}
\epsilon^a_1 = \varepsilon^{\dot{\alpha} \dot a} 
\epsilon^a_{1, \, \dot{\alpha} \dot a}.
\end{equation}
Using any of the equations in (\ref{constraints}) we can determine 
$\epsilon^-_0 = \tau_3\rho^3 \epsilon^-_1 = -\epsilon^-_1$. 

This suggests that those operators can be defined in a 
topologically twisted version of $\cN=4$ SYM. This twisting consists of 
replacing the $SU(2)_R$ factor in the Lorentz group with the sum of 
$SU(2)_R$ and $SU(2)_A$ 
\cite{Yamron:1988qc,Vafa:1994tf}%
\footnote{It is case ii) in \cite{Vafa:1994tf}. 
The loops of \cite{Zarembo:2002an} are related to case i), 
see \cite{Dymarsky:2006ve}.}%
. After the twisting the supercharges decompose under 
the three remaining $SU(2)$ factors (the middle one is the twisted group) as
\begin{equation}
 (\bf{2},\bf{1},\bf{2},\bf{2}) + (\bf{1},\bf{2},\bf{2},\bf{2}) 
\rightarrow 
(\bf{2},\bf{2},\bf{2}) + (\bf{1},\bf{3},\bf{2}) + 
(\bf{1},\bf{1},\bf{2}).  
\end{equation}
From the above it is clear that the two supercharges that annihilate our 
Wilson loops 
are in the $(\bf{1},\bf{1},\bf{2})$, so they become scalars 
after the twisting. As usual, one would then like to regard them as BRST 
charges, and the Wilson loops will be observables in their cohomology. 

Note though that in our case the charges consist of a combination of 
$\bar Q$ and $\bar S$
\begin{equation}
\bar\cQ^a= \varepsilon^{\dot{\alpha} \dot a} 
\left ( \bar{Q}^a_{\dot{\alpha} \dot a} 
- \bar{S}^a_{\dot{\alpha} \dot a} \right),
\label{supercharges}
\end{equation}
so their anti-commutators do not vanish, but 
close on the $SU(2)_B$ generators, which complicates somewhat their 
identification with BRST charges.

To illustrate the richness of this construction we present now seven 
different subclasses of operators which preserve more supersymmetries. 
Each example has some special features that we point out here and will 
elaborate on in \cite{next}.
\begin{enumerate}
\item
{\bf Large $S^2$:}
If we restrict the loop to lie on an $S^2$ defined by, 
say, $x^4=0$, then the left and right forms are no longer independent, 
rather
\begin{equation}
\sigma^L_i=-\sigma^R_i=-2\epsilon_{ijk}x^j\,d x^k\,.
\end{equation}
Then (\ref{susy}) has more solutions. In addition to 
the anti-chiral supersymmetries written above, such a Wilson loop 
also preserves two chiral supersymmetries. 
The generic Wilson loop on $S^2$ will therefore 
preserve $1/8$ of the supersymmetries. 

There is an interesting property of the loops on $S^2$ involving the 
replacement of the gauge and scalar couplings. 
Consider an arbitrary smooth curve on $S^2$ which is 
nowhere a geodesic parameterised by 
$\vec x(s)$, and let us take $|\dot x|=1$. The scalar couplings will 
be given by the standard cross product in three dimensions as 
$\vec\Theta(s)=\dot{\vec x}\times\vec x$. Those 
are also unit vectors in $\bR^3$, so we can 
consider also a loop whose shape is given by $\vec\Theta$. 
A simple calculation shows that the
scalar couplings for the new loop will be proportional to $\vec x$.

This suggests the existence of a duality between the scalar and vector couplings 
and it is tempting to speculate that it will extend to a duality between 
the embedding of the string in the dual description into the $AdS_5$ and 
$S^5$ parts of the geometry.
 
\item
{\bf Large circle:}
By this construction a maximal circle will couple only to a single 
scalar, for example a circle in the $(1,2)$ plane will couple only to 
$\Phi^3$. Studying the supersymmetry variation leads to the single constraint
\begin{equation}
\rho^3\gamma^5 \epsilon_0 = i \gamma_{12} \epsilon_1,
\label{half-constraints}
\end{equation} 
so the loop preserves $16$ ($8$ chiral and $8$ anti-chiral) supercharges.
This is the most studied $1/2$ BPS circular Wilson loop, 
whose perturbative expansion seems to be captured by a Gaussian matrix 
model \cite{Erickson:2000af,Drukker:2000rr}.

\item
{\bf Latitude:}
Consider a non-maximal circle (a latitude) parametrized by 
\begin{equation}
x^{\mu} = (\sin\theta_0 \cos t ,\, \sin\theta_0 \sin t ,\, 
\cos\theta_0,\,  0) \,. 
\label{quarter-circle}
\end{equation}
This is essentially the same Wilson loop operator considered in 
\cite{Drukker:2005cu, Drukker:2006ga}, 
except that by a conformal transformation 
we moved the circle from the equator to a parallel%
\footnote{Also compared to \cite{Drukker:2006ga} $\theta_0$ 
is replaced here by $\pi/2-\theta_0$.}. Here the scalar couplings 
also follow a latitude on the ``dual'' $S^2$, but at $\pi/2-\theta_0$ 
instead of $\theta_0$.

If $\cos\theta_0 \ne 0$, the loop preserves $1/4$ of the supersymmetries. 
This loop also seems to be given by a Gaussian matrix model 
with the only modification that the coupling $g^2$ is 
replaced by $g^2\sin^2\theta_0$.

\item
{\bf Two longitudes:}
Inside a large $S^2$ consider a loop made of two arcs of length $\pi$ 
connected at an arbitrary angle $\delta$, i.e. two longitudes on the 
sphere. We can parametrize the loop in the following way
\begin{equation}
\begin{array}{ll}
x^{\mu}=(\sin t,\, 0,\, \cos t,\, 0)\,,\qquad
& 0 \le t \le \pi\,, \\
x^{\mu}=( -\cos \delta \sin t,\, -\sin \delta \sin t,\,\cos t,\,0)\,,\qquad  
&\pi \le t \le 2 \pi\,.
\end{array}
\label{arcs}
\end{equation} 

The corresponding Wilson loop operator will couple to $\Phi^2$ along the 
first arc and to $- \Phi^2 \cos \delta + \Phi^1 \sin \delta$ along the 
second one. Each arc, being (half) a maximal circle, produces a single 
constraint and is $1/2$ BPS. Together the combined system is 
$1/4$ BPS.

This example has many interesting features. By a stereographic projection 
it is mapped to a cusp in the plane, where along each of the rays the 
scalar coupling is constant. This is an operator of the class constructed 
in \cite{Zarembo:2002an} and has trivial expectation value. 
The observable on $S^2$ is not trivial, rather \cite{next}
\begin{equation}
W\simeq
\begin{cases}
1+\frac{g^2N}{8\pi^2}\,\delta(2\pi-\delta)\,,&g^2N\ll1\,,\\
\exp\sqrt{\frac{g^2N}{\pi^2}\,\delta(2\pi-\delta)}\,,&
g^2N\gg1\,,\\
\end{cases}
\end{equation}
Note that as in the latitude case, the only modification from the 
circle at $\delta=\pi$ is the rescaling of the coupling both at weak 
and at strong coupling by the same factor. But here the perturbative 
calculation does not seem as simple as before.

\item
{\bf Hopf fibers:}
The large circle in the $(1,2)$ plane, mentioned above, coupled only 
to the scalar $\Phi^3$. There is actually a 2-parameter family of circles 
that will couple to this same scalar. They are the integral curves of 
the vector field $\xi_3^R$ dual to $\sigma_3^R$. Each of those 
circles is a fiber in the Hopf-fibration of $S^3$, where the base is 
an $S^2$.

While a single circle preserve 16 supercharges, the combined system 
of two or more fibers will break all the chiral supercharges and will 
preserve the 8 anti-chiral ones. So this system too is $1/4$ BPS. If 
one considers this system in perturbation theory, the effective 
propagator, including both the gauge field and scalar exchange, between 
two arbitrary points along two of those circles will be a constant, 
independent of the distance between the circles. This suggests that 
this system is also described by the Gaussian matrix model, and that 
there is a ``no-force'' condition between two circles when they are moved 
parallel to the Hopf-fibration.

\item
{\bf Anti-chiral $1/8$ BPS loops:}
In a similar way to the last example, where adding more circles broke the 
chiral supersymmetry it is possible to break the chiral supersymmetries 
preserved by the $1/4$ BPS longitudes example. There is a family 
of arbitrary curves (except for one constraint) on the base of the 
Hopf-fibration. The idea is straight-forward, but writing down the 
details is a bit complicated, so we will leave it to \cite{next}.

\item
{\bf Infinitesimal loops:}
Finally, if a loop is very small, concentrated entirely near a single point, 
say $x^4=1$, one will 
not see the curvature of the sphere anymore. The left and right forms 
will then become exact differentials
\begin{equation}
\sigma_i^{R,L}\sim 2 dx_i\,,\qquad
i=1,2,3\,.
\end{equation}
So the Wilson loops (\ref{susy-loop}) reduces to the ones constructed 
by Zarembo in \cite{Zarembo:2002an} (note that this construction 
does not allow for an arbitrary curve in $\bR^4$, but only in $\bR^3$).

This may explain why in this case the expectation value of the loops 
vanishes. The planar loops come from infinitesimal ones on $S^3$, 
so it is quite natural that their VEVs vanish.
\end{enumerate}

We have presented a new class of supersymmetric Wilson loops in 
$\cN=4$ supersymmetric YM. All the examples above, except for 
the large circle, the latitudes and the infinitesimal curves were not 
known before. Those operators will generally have non-vanishing 
expectation values, providing a wide arena for possible calculations 
both on the gauge theory side and in string theory and may lead 
to further tests of the $AdS$/CFT correspondence, as well as 
being perhaps related to topological gauge theory. In addition such 
loops may have 
applications in theories with less supersymmetry.

More details on this construction as well as explicit gauge theory 
and string theory calculations will be provided in an upcoming 
publication \cite{next}.

\subsection*{Acknowledgments}
We would like to thank N. Itzhaki, Y. Oz, J. Plefka, M. Staudacher 
for interesting discussions. 
N.D. is grateful to Tel Aviv University, the Hebrew University and 
the University of Barcelona for their hospitality during the course 
of this work. 
S.G. and D.T. acknowledge partial financial support through the NSF 
award PHY-0354776.

\end{document}